\documentclass[showpacs,amsmath,twocolumn,showkeys,pra]{revtex4-1}
\usepackage{dcolumn}    
\usepackage{bm}         
\usepackage{ifpdf}
\usepackage{amssymb,lineno,amsfonts}
\usepackage{graphicx}   
\usepackage{bbm}
\usepackage{mathrsfs}
\usepackage{upgreek}
\usepackage{epstopdf}
\usepackage{setspace}
\usepackage{hyperref}
\usepackage[matrix,frame,arrow]{xypic}
\usepackage{float}
\usepackage{natbib}
\usepackage{color} 
\usepackage{url}
\usepackage[super,negative]{nth}
\definecolor{med-blue}{RGB}{25,25,112} 
\hypersetup{colorlinks, linkcolor={red},citecolor={blue}, urlcolor={blue}}

\newcommand{\ket}[1]{\vert{#1}\rangle}

\newcommand{\inpr}[2]{\langle{#1}\vert{#2}\rangle}

\begin{document}
	
\title{Rapid Exponentiation using Discrete Operators:\\ Applications in Optimizing Quantum Controls and Simulating Quantum Dynamics}
	\author{Gaurav Bhole and T. S. Mahesh
	}
	\email{mahesh.ts@iiserpune.ac.in}
	\affiliation{Department of Physics and NMR Research Center,\\
		Indian Institute of Science Education and Research, Pune 411008, India}
	
	\begin{abstract}
		{Matrix exponentiation (ME) is widely used in various fields of science and engineering. For example, the unitary dynamics of quantum systems is described by exponentiation of Hamiltonian operators.  However, despite a significant attention, the numerical evaluation of ME remains computationally expensive, particularly for large dimensions.  Often this process becomes a bottleneck in algorithms requiring iterative evaluation of ME.  Here we propose a method for approximating ME of a single operator with a bounded coefficient into a product of certain discrete operators.
        This approach, which we refer to as Rapid Exponentiation using Discrete Operators (REDO), is particularly efficient for iterating ME over large numbers.  We describe REDO in the context of a quantum system with a constant as well as a time-dependent Hamiltonian, although in principle, it can be adapted in a more general setting.
        As concrete examples, we choose two applications.  First, we incorporate REDO in optimal quantum control algorithms and report a speed-up of several folds over a wide range of system size.  Secondly, we propose REDO for numerical simulations of quantum dynamics.  In particular, we study exotic quantum freezing with noisy drive fields.
%
%
		}
	\end{abstract}
	
	\maketitle
	
	\section{Introduction}
Many processes in nature are described by linear differential equations of the form $\dot{y} = Ay$, whose solution is of the form $y(t) = y(0) e^{At}$.  If $A$ is a matrix, then it is necessary to efficiently evaluate the matrix exponential (ME) $e^{At}$.  We look into the situations requiring repeated evaluations of ME, and as examples we describe optimization of quantum controls and simulation of quantum dynamics.
	
``Given a quantum system, how best can we control its dynamics?" is the central question of quantum control. This control over the dynamics of a quantum system is often achieved via  electromagnetic pulses \cite{gordon1997active}. Quantum control had earlier been used to tailor the final output of chemical reactions \cite{shapiro1997quantum,tannor1985control,tannor1986coherent,brumer1986control,peirce1988optimal,kosloff1989wavepacket}. Since then, quantum control has been applied to a wide variety of fields ranging from biology \cite{prokhorenko2006coherent,wohlleben2005coherent}, spectroscopy \cite{vandersypen2005nmr,khaneja2005optimal,maximov2008optimal}, to imaging and sensing \cite{wiki:Coherent_control}. More recently, quantum control has been extensively used in the field of quantum information processing. Information processing devices based on quantum mechanical systems, known as quantum processors, can exploit quantum superpositions and thereby have the ability to outperform their classical counterparts.  The quantum processors are prone to more types of errors and in this sense lack the robustness of classical digital logic.  However, at the same time they also allow intricate quantum operations that can take short-cuts leading to enhanced computational efficiency \cite{nielsen2002quantum}.
Thus, one needs to achieve an efficient control over the dynamics of the quantum device to prove its computational advantage. 

A remarkable progress has been achieved to attain control over the quantum systems during the past several years \cite{chu2002cold,chiorescu2003coherent,collin2004nmr,negrevergne2006benchmarking,ladd2010quantum,o2010quantum}. Although analytical methods have been successful for the control of one or two qubits, one needs to employ numerical optimization algorithms for controlling larger number of qubits. Several algorithms (e.g.  \cite{SMP,khaneja2005optimal,maximov2008optimal,de2011second,crab,grivopoulos2003lyapunov,hou2012optimal,lyapunov,bhole2016steering,lucarelli2016quantum}) for quantum control have been proposed in the past years.

The central part in many of the quantum control algorithms as well as in simulating quantum dynamics is generating the unitaries by exponentiating Hamiltonians, and this part is often iterated over large numbers.  ME is a very important problem in theoretical computer science and accordingly  numerous ways have been explored for this task \cite{moler1978nineteen,moler2003nineteen}.
 These include polynomial methods, series methods, differential equations method, matrix decomposition method, etc.
Since the matrix-exponentiation forms a bottle-neck in the quantum control algorithms as well as in other  simulations of quantum dynamics, any improvement in speeding up this process can greatly improve the efficiency of these algorithms.

In this article, we describe the method, Rapid Exponentiation using Discrete Operators (REDO), that overcomes the above bottleneck by keeping ME outside the iterations.  In the following section we provide the theory of REDO.  In section III and IV we describe the applications of REDO in quantum control and in simulating quantum dynamics respectively.   Finally we conclude in section V.

\section{Rapid Exponentiation by using Discrete Operators (REDO):}
Consider a quantum system governed  by the Hamiltonian 
\begin{eqnarray}
\mathcal{H}_\mathrm{tot}(t) = \mathcal{H}_0+\mathcal{H}(t),
\end{eqnarray}
where $\mathcal{H}_0$ and $\mathcal{H}(t)$ are fixed and  controllable parts respectively. 
Let $\ket{\psi(t)}$ be the state vector in the Schr\"{o}dinger representation that evolves according to $\ket{\psi(t)} = U(t,t_0)\ket{\psi(t_0)}$ through the propagator
\begin{eqnarray}
U(t,t_0) = 
\mathcal{D}\left[\exp\left(
-i\int_{t_0}^t \mathcal{H}_\mathrm{tot}(t) dt \right)\right],
\label{UHtot}
\end{eqnarray}
where $\mathcal{D}$ is the Dyson time-ordering operator.  
The explicit influence of the fixed part of the Hamiltonian can be eliminated by going into Dirac representation via the transformation $V^\dagger(t) = e^{i\mathcal{H}_0 t}$.
The state vectors in the Schr\"odinger $(\ket{\psi(t)})$ and the Dirac $(\ket{\widetilde{\psi}(t)})$ representations are related by
\begin{eqnarray}
\ket{\widetilde{\psi}(t)} &=& V^\dagger(t)\ket{\psi(t)}
\nonumber \\
&=& V^\dagger(t) U(t,t_0)\ket{\psi(t_0)}
\nonumber \\
&=& V^\dagger(t) U(t,t_0)V(t_0)\ket{\widetilde{\psi}(t_0)},
\label{u0t}
\end{eqnarray}
from which, we obtain the Dirac propagator
\begin{eqnarray}
\widetilde{U}(t,t_0) = V^\dagger(t) U(t,t_0)V(t_0).
\label{u2ud}
\end{eqnarray}
Differentiating the above equation w.r.t. time leads to the Schr\"{o}dinger equation in Dirac representation  
\begin{eqnarray}
\frac{d}{dt}{\widetilde{U}(t,t_0)} &=&
-i\widetilde{\mathcal{H}}(t) \widetilde{U}(t,t_0),
\end{eqnarray}
where $\widetilde{\mathcal{H}}(t) =  V^\dagger(t)\mathcal{H}(t)V(t)$ is the
control Hamiltonian in the Dirac representation. Solution of the above equation is of the form
\begin{eqnarray}
\widetilde{U}(t,t_0) = \mathcal{D}\left[\exp\left(
-i\int_{t_0}^t \widetilde{\mathcal{H}}(t) dt \right)\right].
\end{eqnarray}

In view of the mathematical simplicity as well as   control-implementation by a digital hardware, we discretize time into $N$ equal segments each of duration $\Delta t$, thus leading to piecewise-constant control Hamiltonian.  The discrete Hamiltonians in the Dirac representation are of the form 
\begin{eqnarray}
 \widetilde{\mathcal{H}}_n = \Omega_n \widetilde{\cal{S}},
 \end{eqnarray}
where $\widetilde{\cal{S}}$ is a fixed operator and $\Omega_n \in [0,\Omega_\mathrm{max}]$ is the non-negative control parameter.  Then the propagator for this time interval simplifies to
 \begin{eqnarray}
\widetilde{U}_n = \exp\left({-i \Omega_n \Delta t  \widetilde{\cal{S}}}\right).
\label{untilde}
\end{eqnarray}
Given a large set $\{\Omega_n\}$, the objective is to efficiently evaluate the propagators $\{\widetilde{U}_n\}$.

We denote $\mathbb{N}_b = \{0,1,\dots,b-1\}$ as the set of  integers where $b \ge 2$.  
We first coarse grain the scaling parameter such that
\begin{eqnarray}
\Omega_n &=& \sum_{j=l}^{m} c_{nj} b^{j}  + \mathcal{O}(\epsilon), \nonumber \\
&=& \lfloor \Omega_n \rceil + \mathcal{O}(\epsilon),
\label{cgrain}
\end{eqnarray}
where the coefficient $c_{nj} \in \mathbb{N}_b$.
The lower-limit $l$ is chosen sufficiently small to meet the precision of graining, i.e., $b^{l} = \epsilon$.  On the other hand, the upper-limit $m$ is chosen to cover the range of $\Omega_n$, i.e., $b^{m+1} > \Omega_\mathrm{max}$.  Thus,
\begin{eqnarray}
l &\simeq& \log_b(\epsilon) 
~ \mbox{and,} \nonumber \\
m &\simeq& \log_b (\Omega_\mathrm{max})-1.
\end{eqnarray} 

For example, given $\epsilon = 1$ rad/s and $\Omega_\mathrm{max} = 2.6 \times 10^5$ rad/s, we may choose, $l=0$, $b = 64$, and $m=2$. Thus the course graining leads to discrete amplitudes \begin{eqnarray}
\lfloor \Omega_n \rceil = c_{n0} + c_{n1}64 + c_{n2}64^2,
\end{eqnarray}
so that
\begin{eqnarray}
\widetilde{U}_n \approx 
e^{-ic_{n0}\Delta t\widetilde{S}} \cdot
e^{-ic_{n1} 64 \Delta t\widetilde{S}} \cdot
e^{-ic_{n2} 64^2 \Delta t\widetilde{S}}. 
\end{eqnarray}
Since $c_{nj}$ can take values between 0 and 63, there are only 
$63 \times 3$ (ignoring identity operators) discrete operators of the form in the rhs of above equation.  If these 189 operators are evaluated one-time and stored, the propagator for any $\lfloor \Omega_n \rceil \in [0,64^3]$ rad/s can be obtained simply by at most two matrix multiplications.

REDO algorithms involves three processes. 
\begin{enumerate}
	\item 
A one-time process to evaluate and store $s=(b-1)(m-l+1)$ discrete operators
\begin{eqnarray}
\widetilde{u}_{nj} = e^{-i c_{nj} b^{-j} \widetilde{\cal{S}}}.
\label{redoform}
\end{eqnarray}
\item
The second step is the matrix multiplication 
\begin{eqnarray}
 \widetilde{U}_n \approx \prod_{j=l}^m \widetilde{u}_{nj}.
\label{exp2prod}
\end{eqnarray}
Thus, the problem of evaluating matrix exponentials is reduced into that of $p= (m-l)$ matrix multiplications.  In principle, such processes can also be highly parallelized \cite{waldherr2010fast,schulte2012fast}. 
\item
In the coarse-grained set $\{\lfloor \Omega_n \rceil\}$ each of the repeated entries must be identified and evaluated only once.
\end{enumerate}

\begin{figure}
	\centering
	\includegraphics[trim=1cm 6.5cm 0.5cm 6.5cm, clip=true,width=7cm]{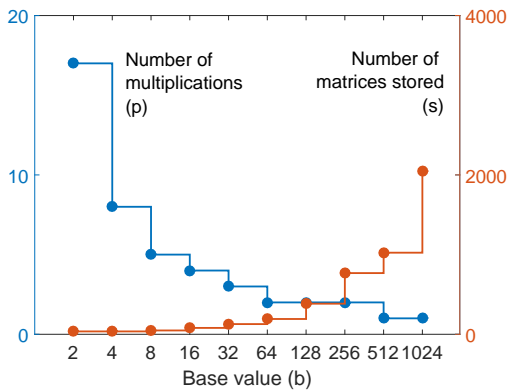}
	\caption{Number of multiplications and required storage versus the base number for $\alpha_\mathrm{max}/\epsilon = 64^3$.}
	\label{sp}
\end{figure}

The computational costs can be quantified in terms of the number of matrix multiplications 
\begin{eqnarray}
p = m-l \simeq \log_b\left(\frac{\Omega_\mathrm{max}}{b\epsilon}\right), ~~
\end{eqnarray}
as well as the number of matrices to be stored 
\begin{eqnarray}
s  =  (b-1)(p+1).
\end{eqnarray}
Fig. \ref{sp} displays profiles of $p$ and $s$ for a range of base values and $\lambda$ values.
Thus depending on the computational resource, an optimal base number $b$ is to be chosen that balances the storage $s$ and the number of matrix multiplications $p$ which maximizes the overall computational efficiency.

\begin{figure}[b]
	\centering
	\includegraphics[trim=1cm 1cm 2cm 1cm, clip=true,width=9cm]{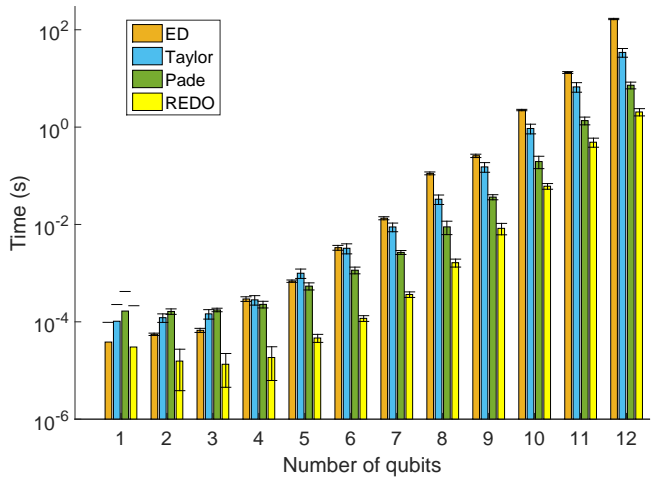}
	\caption{Processing time per propagator using eigen-decomposition (ED), Taylor series, Pade, and REDO algorithms for different system sizes.  Here the following values are used for various parameters: $\Omega_\mathrm{max} = 2.6 \times 10^5$ rad/s, $\epsilon = 1$ rad/s, $b=64$, $l=0$, and $m=2$.}
	\label{comp1}
\end{figure}

The processing time per propagator using various numerical methods (implemented in MATLAB) for different system sizes are compared in Fig. \ref{comp1}.  Evidently, REDO achieves a speed-up of 3 to 10 times over the next fastest method, i.e., Pade algorithm.

\begin{figure}
	\centering
	\includegraphics[trim=1cm 2cm 1cm 2cm, clip=true,width=7cm]{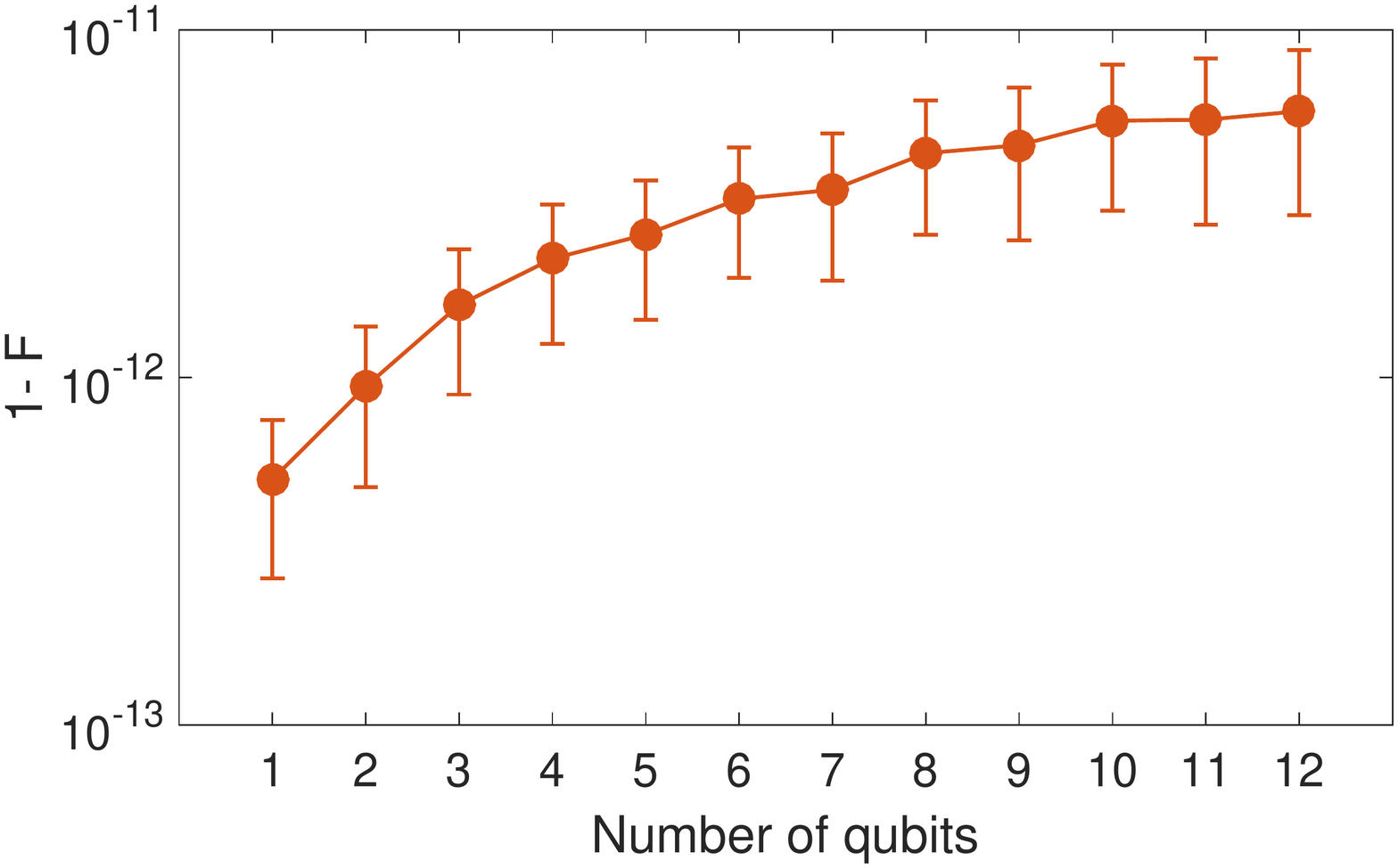}
	\caption{Deviation between the propagators evaluated by REDO algorithm and Pade algorithm for $\epsilon = 1$ and $\Delta t = 5~ \upmu$s.}
	\label{deviation}
\end{figure}

The overlap between a propagator and its coarse grained version is given by the fidelity
\begin{eqnarray}
F &=& \left| \mathrm{Tr}
\left[
e^{i \Omega_n \Delta t  \widetilde{\cal{S}}} \cdot
e^{-i \lfloor \Omega_n \rceil  \Delta t  \widetilde{\cal{S}}}
\right]/N
\right| ^2 \nonumber \\
&=& \left| \mathrm{Tr}
\left[e^{i \mathcal{O}(\epsilon) \Delta t \widetilde{\cal{S}}}
\right]/N
\right| ^2,
\end{eqnarray}
where we have used Eq. \ref{cgrain}.
In our case, the operator $\widetilde{\cal{S}}$ is  traceless and involutory, and therefore the deviation
\begin{eqnarray}
1 - F \approx \mathcal{O}(\epsilon^2 \Delta t^2).
\end{eqnarray}
Thus REDO algorithm with sufficiently small $\epsilon$ and $\Delta t$ can lead to speed-up without any significant loss of fidelity.
The deviations for propagators evaluated by REDO algorithm against those by Pade algorithm for various system size are shown in Fig. \ref{deviation}.

In the following sections we explain how the above procedure can be applied in quantum control as well as in simulating quantum dynamics.

\section{REDO for quantum control}
Although the methods described can be adapted to a general quantum system, we have chosen the NMR setting for the sake of clarity.
Consider an NMR system having resonance offsets $\omega_i$, indirect spin-spin interactions $J_{ij}$, and direct spin-spin interactions $D_{ij}$.
The total Hamiltonian is 
$\mathcal{H}_\mathrm{tot}(t) = \mathcal{H}_0 + \mathcal{H}(t)$. The internal Hamiltonian is of the form
\begin{eqnarray}
{\cal H}_0 = -\sum_i \omega_i I_{iz}
+ 2\pi \sum_{i<j} J_{ij}  \textbf{I}_{i}\cdot\textbf{I}_{j} 
\nonumber \\
+ 2\pi \sum_{i<j}D_{ij}(3I_{iz}I_{jz}-\textbf{I}_{i}\cdot\textbf{I}_{j}),
\end{eqnarray}  
where $I_{iz}$ etc. are the components of spin angular momentum operators $\textbf{I}_i$.
The control Hamiltonian involves the time-dependent terms, i.e.,
\begin{eqnarray}
\mathcal{H}(t) &=& \sum_{k} \Omega_k(t) 
\left[
S_{kx}\cos\phi_k(t) + S_{ky}\sin\phi_k(t)
\right],
\end{eqnarray}
with $\Omega_k(t)$ being the RF amplitude on the $k$th nuclear species with a collective spin operator $\textbf{S}_k = \sum_j \textbf{I}_{j(k)}$, where the sum is carried over only the spins of $k$th nuclear species. For the sake of simplicity, one normally considers piecewise-constant control parameters such that   the evolution of the quantum system is described by the propagator
\begin{eqnarray}
U = U_N U_{N-1} \dots U_2 U_1,
\end{eqnarray}
where 
\begin{eqnarray}
U_n = \exp\left[-i \Delta t \left({\mathcal{H}_0}+\mathcal{H}_n \right)\right],
\label{propUn}
\end{eqnarray}
where $\Delta t = T/N$ is the duration of each of the $N$ discretized segments over a total duration $T$.  The discretized control Hamiltonian corresponds to that of a linearly polarized RF wave in xy plane, i.e., 
\begin{eqnarray}
\mathcal{H}_n &=& \sum_k H_{kn} \nonumber \\
&=& \sum_{k} \Omega_{kn} 
\left[
S_{xk}\cos\phi_{kn} + S_{yk}\sin\phi_{kn}
\right] \nonumber \\
&=& Z_{n} \left(
\sum_{k}  \Omega_{kn} S_{xk}
\right)
Z^\dagger_{n},
\label{vn}
\end{eqnarray}
where $H_{kn}$ denotes the Hamiltonian for each nuclear species and $Z_{n} = 
\exp(-i \sum_k \phi_{kn} S_{kz} )$.  Thus a linear polarized wave about an arbitrary axis in the xy plane is treated as an x-polarized wave rotated about z-axis.

The goal is to determine the control profiles $\{\Omega_{kn}, \phi_{kn}\}$ which generate a desired propagator $U_f$ by maximizing the unitary-fidelity
\begin{eqnarray}
F_U = \left\vert\frac{\mathrm{Tr}
\left[U_f^\dagger U \right]}{\mathrm{Tr}\left[U_f^\dagger U_f \right]} \right\vert^2
\end{eqnarray}
subject to certain constraints.
Often, the goal is instead to achieve a quantum state transfer $\ket{\psi(0)} \rightarrow \ket{\psi(T)}$ by maximizing the state-fidelity
\begin{eqnarray}
F_S = \vert \inpr{\psi_f}{\psi(T)} \vert^2,
\end{eqnarray}
where $\ket{\psi_f}$ is the desired state.

These tasks are generally achieved  by gradient methods \cite{khaneja2005optimal} or by global optimization methods \cite{maximov2008optimal}
.  The most time-consuming subroutine in all these methods is the repeated ME described in Eq. \ref{propUn} for the evaluation of propagators.  Although a recent proposal of using bang-bang control alleviates this bottleneck, any speed up of evaluating the propagators $\{U_n\}$ is important for other standard methods based on smooth modulations \cite{bhole2016steering}.  In the following we describe how to adapt REDO algorithm for this purpose.

Transforming to the Dirac representation, the control Hamiltonian in Eq. \ref{vn} becomes
\begin{eqnarray}
\widetilde{\mathcal{H}}_n = Z_{n} \left(
\sum_{k}  \Omega_{k,n} \widetilde{S}_{xk}
\right)Z^\dagger_{n},
\label{vnd}
\end{eqnarray} 
where $\widetilde{S}_{xk} = V^\dagger(\Delta t) S_{xk}V(\Delta t)$ are the $x$-operators in Dirac representation.  Here we have also used the fact that $Z_{n}$ remain unchanged under the transformation since they commute with $\mathcal{H}_0$ and therefore with $V(\Delta t)$.
Finally using Eqs. \ref{u2ud} and \ref{untilde} along with Eq. \ref{vnd}, we obtain
\begin{eqnarray}
U_n &=& V(\Delta t)~ \exp(-i \widetilde{\mathcal{H}}_n \Delta t) ~ V^\dagger(0)
\nonumber \\
&=& V(\Delta t) Z_{n} \left(\prod_k\widetilde{X}_{kn}\right)
Z^\dagger_{n},
\end{eqnarray}
where
\begin{eqnarray}
\widetilde{X}_{kn} = \exp \left(
-i  \Omega_{kn} \Delta t \widetilde{S}_{xk}
\right),
\end{eqnarray}
and we have used $V(0) = \mathbbm{1}$.

\begin{figure}
	\centering
	\includegraphics[trim=3cm 0cm 2.5cm 0cm, clip=true,width=8.5cm]{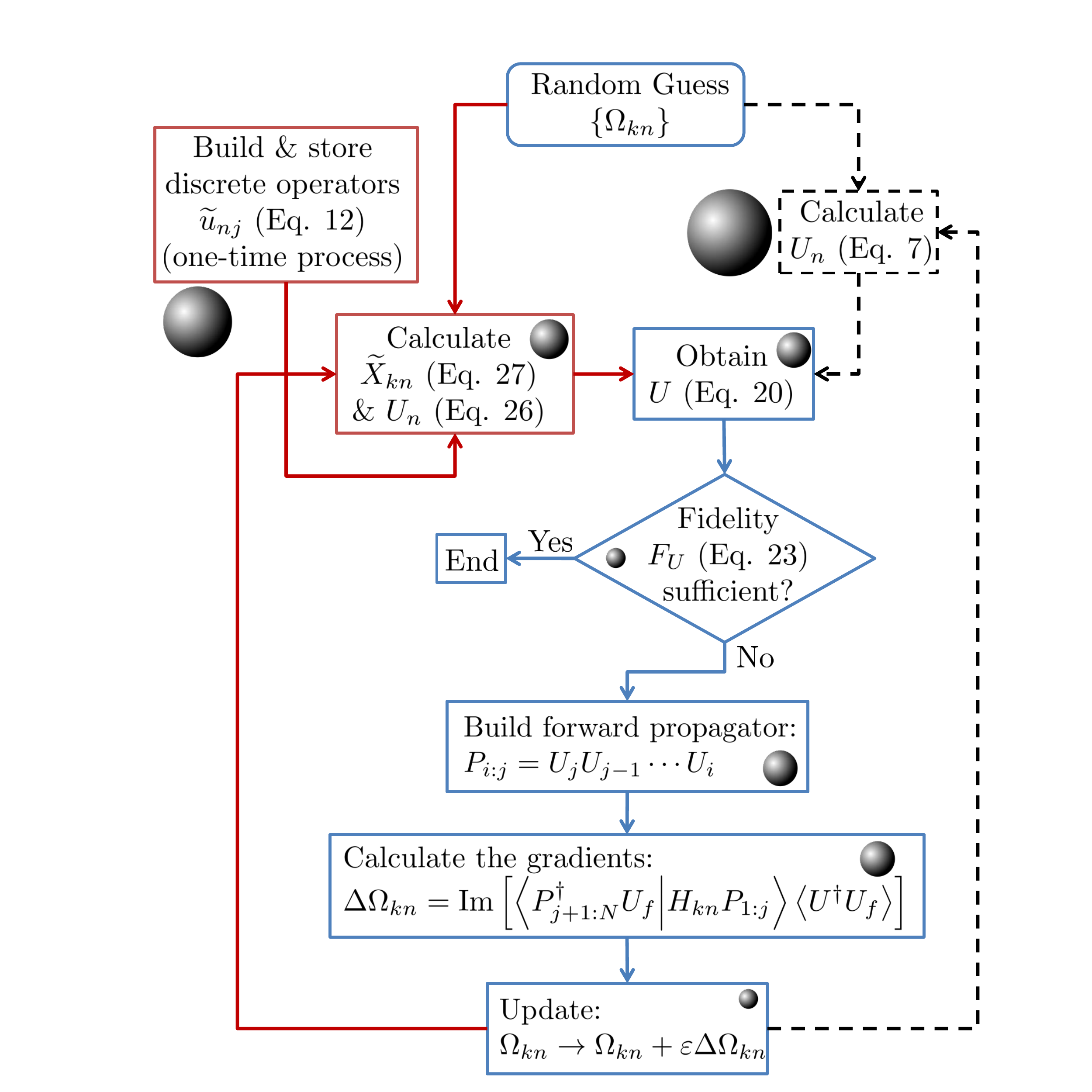}
	\caption{The flowchart describing various processes in REDO-GRAPE algorithm (red outlines) as compared to GRAPE algorithm (dashed outlines). The processes which are common to both are shown in blue lines. The relative computational complexities of various processes are schematically represented by the sizes of  the spheres.}
	\label{fc}
\end{figure}

Since $\widetilde{X}_{kn}$ is of the form of $\widetilde{U}_n$ in Eq. \ref{untilde}, we apply the REDO procedure and convert the matrix exponential into the product as described in the previous section.

Although, REDO can be incorporated in various quantum control algorithms such as GRAPE \cite{khaneja2005optimal}, Krotov \cite{maximov2008optimal}, bang-bang \cite{bhole2016steering}, and Lyapunov \cite{lyapunov}, in the following we describe the GRAPE quantum control algorithm  which is widely used in various experimental architectures \cite{fisher2010optimal,tsai2008gradient,fisher2010optimal}.
For the sake of completeness, we include a brief description of the GRAPE algorithm.  The flow-chart describing various steps of the algorithm are shown in Fig. \ref{fc}.  Being a local search algorithm, GRAPE begins with a set of initial control parameters $\{\Omega_{kn}\}$ that is often chosen randomly.  In order to calculate the updates for each of the control parameters, one needs to evaluate ME of Hamiltonian and obtain the propagators $U_n$.  This is the most expensive step in terms of computational complexity, particularly for large system sizes.  In the flowchart shown in Fig. \ref{fc}, the relative computational complexities in various processes are schematically represented by the sizes of  the spheres. In REDO-GRAPE, the calculation of the discrete operators $\widetilde{u}_{nj}$ is an one-time process that is not part of the iteration loop.

\begin{figure}[t]
	\centering
	\includegraphics[trim=0cm 0cm 0cm 1cm, clip=true,width=9cm]{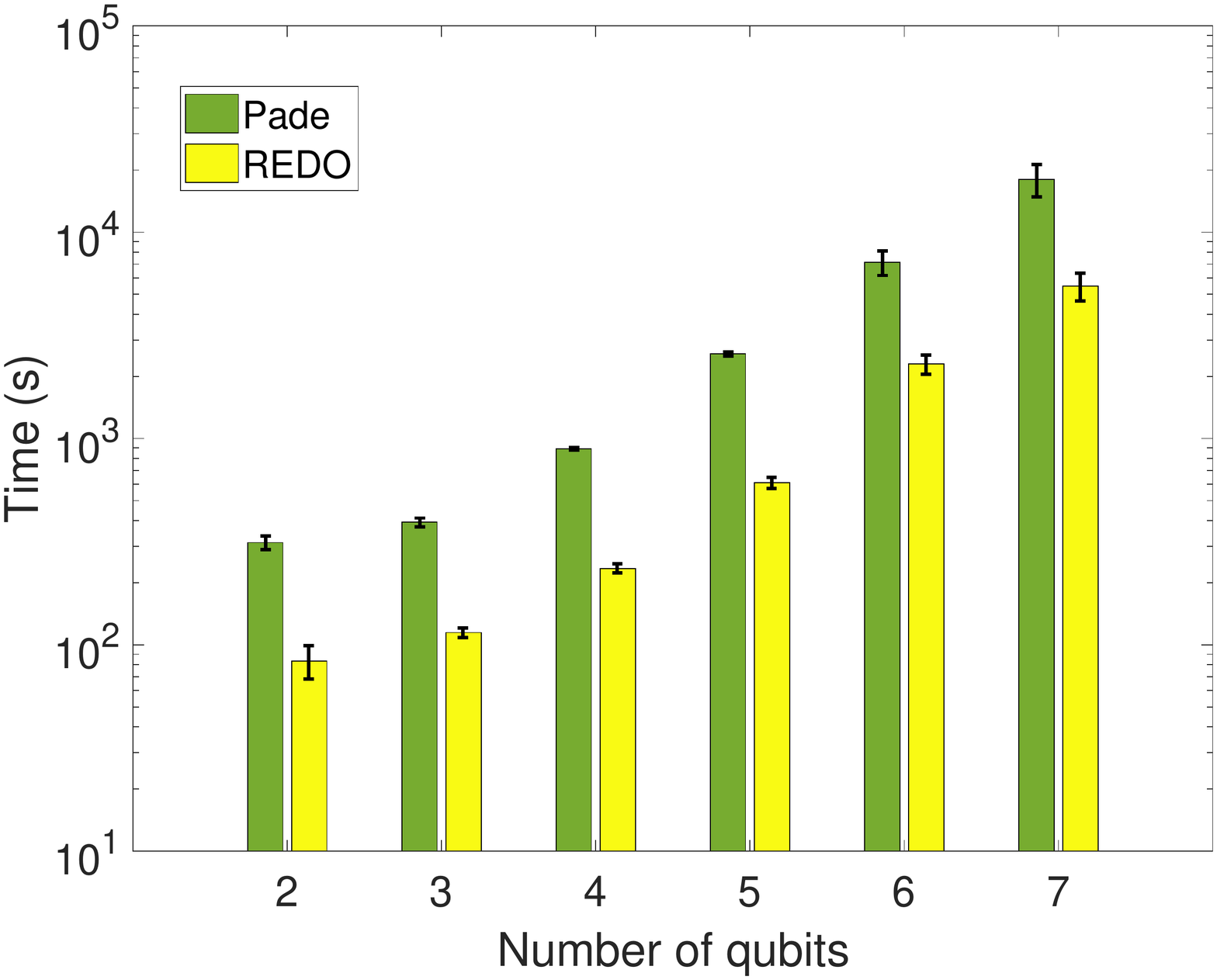}
	\caption{Processing times per iteration for the GRAPE algorithm with Pade as well as REDO method for propagator evaluations.}
	\label{comp}
\end{figure}

Fig. \ref{comp} compares the overall processing times per iteration for the GRAPE algorithm with either Pade or REDO methods for evaluating the propagators with various system sizes.
Clearly, REDO displays a speed-up by at least 3-folds.  For example, REDO-GRAPE with 4-qubits is faster than the usual GRAPE with 2-qubits.

\section{REDO for simulating quantum dynamics}
Consider a quantum system governed by a time-dependent Hamiltonian ${\cal H}_\mathrm{tot}(t)$.  The evolution is described by the unitary $U(t,t_0)$ as in Eq. \ref{UHtot}.  Suppose $M$ is an observable and we are interested in the expectation value $\langle M \rangle = \mathrm{Tr} \left[M \rho(t)\right]$, where $\rho(t) = U(t,t_0) \rho(0) U(t,t_0)^\dagger $. Often numerical simulations would involve iterative ME as illustrated by the following example.  Here we explore the efficiency of REDO method for such a simulation of quantum dynamics.

\begin{figure}[b]
	\centering
	\includegraphics[trim=1cm 0cm 1.5cm 1cm, clip=true,width=8.5cm]{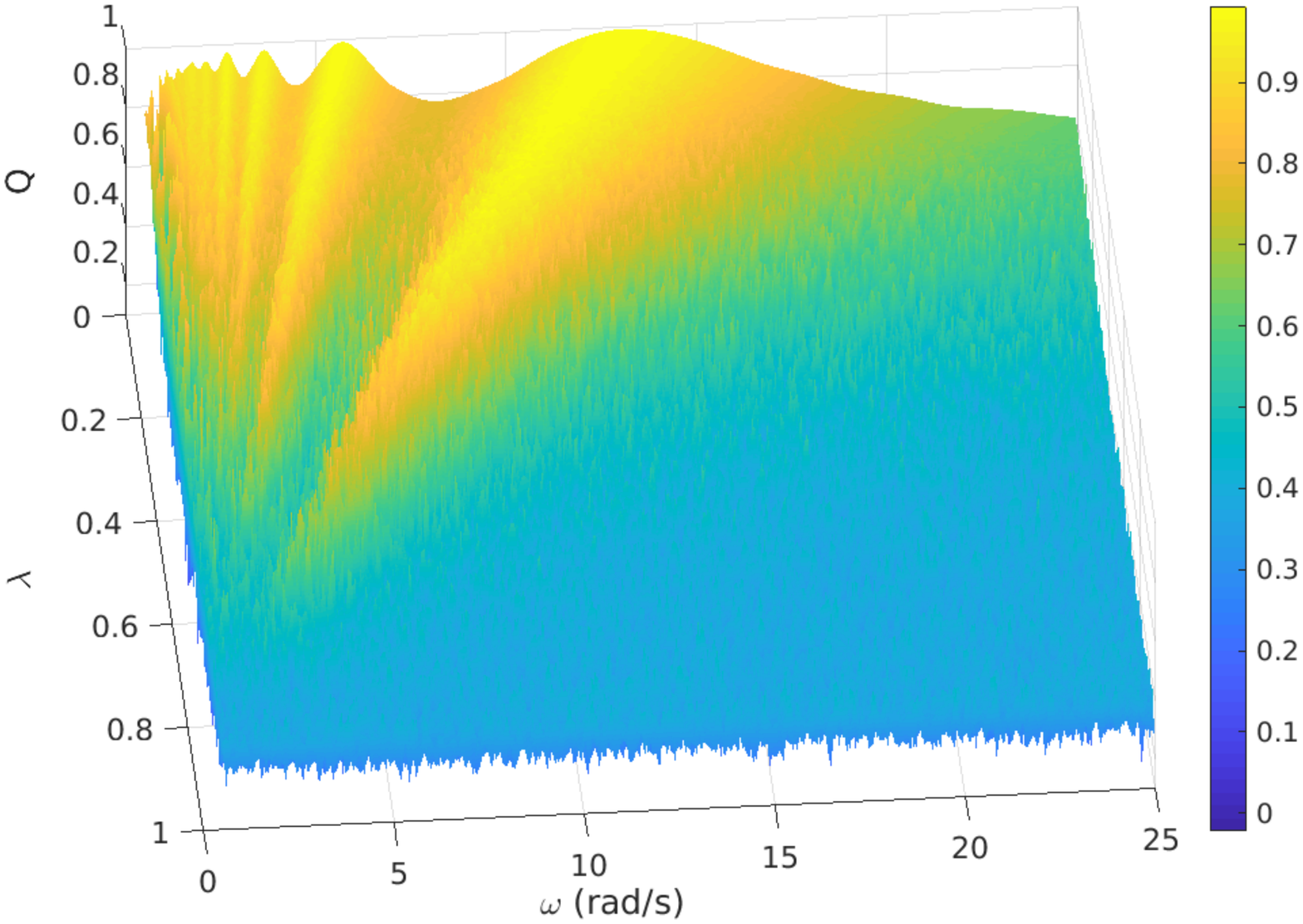}
	\caption{REDO simulation of exotic quantum freezing.  Dynamical order parameter $Q$ is plotted versus drive frequency $\omega$ and the noise parameter $\lambda$.}
	\label{freeze}
\end{figure}

\subsection*{Effect of noisy drive in Quantum Exotic Freezing}
As an example, we simulate an interesting phenomena known as exotic quantum freezing. Classically a driven system freezes its dynamics as the drive frequency gets much higher than its intrinsic frequency.  However, it was shown by Arnab Das that under certain circumstances, a quantum many-body system can show non-monotonic sequence of freezing and nonfreezing behaviour w.r.t. the drive frequency \cite{Arnab}.  The system considered was an Ising chain subjected to a transverse periodic drive of amplitude $h_0$ and frequency $\omega$. The total Hamiltonian is
\begin{eqnarray}
\mathcal{H}_\mathrm{tot}(\omega,t) =  -J  \sum_{i} 2I_{i,z} I_{i+1,z} - 
h_0 \cos \omega t \sum_{i} I_{i,x},~~~~
\end{eqnarray}
where $J$ is the strength of nearest neighbour Ising interaction. Given the initial state $\rho(\omega,0)$, our aim is to find the dynamical order parameter $Q$, i.e., an average response, say that of x-magnetization, over a long duration $T$ (i.e., $t \in [0,T]$).  Theoretically, for an infinite Ising spin chain
\begin{equation}
Q(\omega,T) = \left\langle \mathrm{Tr} \left[\rho(\omega,t) \left(\sum_i I_{i,x}\right)\right] \right \rangle = 
\frac{1}{1+\mathcal{J}_0(2h_0/\omega)},
\end{equation}
where $\rho(\omega,t) = U(\omega,t)\rho(\omega,0)U^\dagger(\omega,t)$ is the instantaneous density matrix of the driven system 
and $\mathcal{J}_0$ is the zeroth order Bessel function \cite{Arnab}.  As the drive frequency $\omega$ is increased from low values, $Q$ reaches unity at various roots of the Bessel function $\mathcal{J}_0(2h_0/\omega)$, indicating non-monotonic freezing behaviour.  

Later Swathi et al have studied this phenomenon experimentally using a three-spin NMR system \cite{swathi}.  They observed that although the decoherence in practical systems reduces the order parameter $Q$, the freezing-frequencies are not shifted.

In the following, we describe the simulation of exotic quantum freezing with a noisy drive by replacing the amplitude $h_0\cos{\omega t}$ with $h_0 \left\{(1-\lambda)\cos{\omega t} + \lambda\eta \right\}$, where the noise parameter $\lambda \in[0,1]$ controls the extent of the random field $\eta \in [-1,1]$ in the drive.

To study quantum exotic freezing under a noisy drive, we consider a linear chain of three spin-1/2 particles with $J=h_0/20$ and $h_0 = 5\pi$.  Taking $\rho(\omega,0) = \sum_{i=1}^{3}I_{i,x}$ as the initial state, we performed numerical simulations by both Pade ME as well as by REDO methods for 500 values of $\omega$ uniformly distributed  from 1 to 25 rad/s and for 1000 values of $\lambda$ uniformly distributed in the range 0 to 1. For each of the $\omega$ values, we performed a long time average for 10,000 time-points uniformly distributed in the range 0 to $20\pi$.
In the REDO method, the transverse coefficient  $\left\{(1-\lambda)\cos{\omega t} + \lambda \eta \right\}$ was coarse grained with $b = 100$, $l=-2$, and $m=-1$ (see Eqn. \ref{cgrain}). We found that the REDO method to be about 6 times faster than repeated Pade ME. 

The simulation result is shown in Fig. \ref{freeze}.  
For $\lambda=0$, it reproduces the result discussed by Swathi et al \cite{swathi}. The points $Q\approx 1$ indicate the freezing regions. As expected, with increasing noise parameter $\lambda$ the freezing gradually disappears indicating completely random dynamics averaging out the response.  However, interestingly, the freezing points tend to move towards lower drive frequencies with increasing noise, presumably due to contributions from high-frequency components of noise.  This phenomenon is yet to be studied experimentally.

\section{Conclusions}
Though matrix exponentiation (ME) appears very often in various branches of science and engineering, it remains a computationally expensive task for large dimensions. Motivated by applications such as optimization of quantum controls and simulation of quantum dynamics, we explored a way of avoiding  iterative evaluations of ME.
We proposed a method for faster evaluation of ME which we referred to as Rapid Exponentiation by Discrete Operators (REDO). This method achieves a substantial speed-up over existing methods of iterative ME.  The REDO overhead includes pre-calculating, storing, and efficiently recycling certain discrete operators. 
We benchmarked REDO against iterative ME with Pade polynomial, eigen-decomposition, and Taylor series, and observed REDO to be faster by several folds over a range of system size.  

The REDO method is general and can be applied in a wide variety of scenarios wherein repeated ME is required. 
In-order to demonstrate the practicality of REDO, we showed its application in quantum control algorithms and to simulate the dynamics of a quantum system. In particular, the REDO method was incorporated in GRAPE algorithm for quantum control and it displayed at least 3-fold speed-up over GRAPE using iterative Pade ME. To demonstrate the superiority of REDO in simulation of quantum systems, we studied and simulated exotic quantum freezing phenomena in a three-spin Ising chain. Here, REDO was outperformed iterative Pade ME by over six times.  It may be possible to apply REDO in many other scenarios, and we expect it to become a standard protocol for handling repeated ME.

\begin{acknowledgments}
This work was partly supported by DST/SJF/PSA-03/2012-13 and CSIR 03(1345)/16/EMR-II. GB acknowledges support from DST-INSPIRE fellowship.
\end{acknowledgments}

\bibliography{ref}{}
\bibliographystyle{apsrev4-1}
\end{document}